*Research Article*

# Optical Network Models and Their Application to Software-Defined Network Management

**Thomas Szyrkowiec,[1] Achim Autenrieth,[1] and Wolfgang Kellerer[2]**

[1]*ADVA Optical Networking SE, Munich, Germany*
[2]*Chair of Communication Networks, Technical University of Munich, Munich, Germany*

Correspondence should be addressed to Thomas Szyrkowiec; tszyrkowiec@advaoptical.com

Software-defined networking is finding its way into optical networks. Here, it promises a simplification and unification of network management for optical networks allowing automation of operational tasks despite the highly diverse and vendor-specific commercial systems and the complexity and analog nature of optical transmission. Common abstractions and interfaces are a fundamental component for software-defined optical networking. Currently, a number of models for optical networks are available. They all claim to provide open and vendor agnostic management of optical equipment. In this work, we survey and compare the most important models and propose an intent interface for creating virtual topologies which is integrated in the existing model ecosystem.

## 1. Introduction

The raise of software-defined networking (SDN) is leading to a paradigm shift in network operations. The separation of control and data plane pushed the development of (logically) centralized controller entities. Those controllers are responsible for the configuration of the data plane. In the case of optical networks, every vendor has his own closed network management system (NMS) with access to the devices through proprietary interfaces. This fact leads to vendor islands in the network. Even though the NMS is a centralized entity, it has a much wider scope than a controller and lacks standardized models and open—southbound and northbound—interfaces so far. Additionally, optical equipment is modular, which leads to node constraints based on the hardware configuration. Unlike current IP-based networks, the optical domain most commonly deploys circuit switching. The assumption that every (cross-) connection between input and output ports is possible, as with packet-switched networks, does not hold for a realistic optical network model that will be adopted by vendors and operators. On a physical network level, the analog nature leads to network constraints; for example, optical signal-to-noise ratio, crosstalk, and dispersion affect the maximum signal reach. They have to be taken into account, before setting up services through the network, and therefore need to be modeled correctly.

5G will introduce flexibility into the next generation of mobile networks. They will require principles of SDN to enable flexible control and automation of network management tasks on that infrastructure. One new concept, introduced by 5G, is the creation of service aware slices [1, 2]. It is expected that particular use cases or applications will receive specialized network slices with agreed on properties. They will be using a common or shared infrastructure, which is capable of employing network virtualization techniques. The software, which is commonly used for virtualizing networks, is called network hypervisor [3]. It represents a mediation layer between the hardware and northbound client controllers. It creates virtual topologies, exposes them to its clients, and translates the incoming control commands appropriately. Since optical transport networks are a prominent part of the core network in current-generation and next-generation mobile network deployments, they need to support a similar functionality [4]. Those optical network hypervisors will be capable of maintaining optical resources and exposing multiple slices to tenants. In the optical domain, new open and standardized interfaces are a requirement, on one hand, to allow applications to request and control network slices and, on the other hand, to unify the control of hardware coming from different vendors.

A variety of optical network models, covering various aspects, have been published in recent years. They range from generic descriptions of optical network elements, in order to unify the configuration process, to network controller orchestration models, to allow the management of network-wide services [5]. One of the factors leading to this increasing number of models is the modelling language YANG [6]. YANG facilitates the creation of models that can be easily applied to protocols [7] as well as to code generation. All models have the goal of establishing an open unified network management for the base functionality of optical equipment, independent of the vendor, instead of using closed NMSs with proprietary interfaces. The applicability of SDN to optical networks relies on such configuration and topology models in order to offer controllers for the optical domain or to include optical networks in a multidomain SDN environment. For this reason, existing management models are ported to YANG, for example, ITU-T G.874.1 (11/2016), and undergo a unification process now; that is, they are based on the ONF core model [8] and follow the same modelling guidelines. Operators expect to optimize their operations through unified models, assuming that controllers with a global view simplify network management and make it more dynamic and efficient through automation.

Reference [9] comprehensively surveys techniques applying SDN to optical networks including virtualization and orchestration. It concludes that simplified management strategies are required but does not cover the models for a centralized optical network management. Our work aims at closing this gap. We survey the most relevant models defined by standardization bodies and open industry alliances and characterize them. We focus on specialized models that are trying to capture properties of optical networks and abstract them to a set of parameters—the least common denominator. We give an overview of the established and most mature ones that are currently available. In addition, we identify virtual topology creation as a gap in the currently available descriptions and propose a new model for closing it.

The work is structured as follows. First, we introduce a common terminology. In Section 3, we list a set of network management functions that represent application areas for the models. Then we introduce and evaluate the selected optical models. Finally, we present the intent model for virtual topology creation in Section 6.

## 2. Common Terminology

We provide some general definitions of commonly used terms, which are mapped to more specific definitions in the model comparison.

A graph is defined as a set of vertices interconnected by edges. A network topology is a graph containing nodes (vertices) and links (edges). For the models, links are assumed unidirectional, if not stated otherwise. A generic node contains a number of link termination points (TPs). Link TPs, also referred to as endpoints, represent attachment points for links. In general, they are not assigned to any particular layer, even though some models redefine them for special purposes and assign them to layers. Links represent an adjacency between two link TPs of two distinct nodes. A path lists the traversed nodes and links in a sorted order. The exact definition differs in the presented models, for example, the type of the path's endpoints. Virtual topologies, sometimes called abstracted or customized, refer to a network view, which is different from the underlying physical network topology. This is typically achieved by hiding details of the underlying network. A network slice is a portion of a network, similar to a virtual topology, which is created according to a set of requirements, for example, capacity, latency, and availability. The difference is that the slice is provisioned from the start and used by services sharing common characteristics.

## 3. Network Management Functions

We have identified a set of application areas for optical models based on an infrastructure provider's view. A provider managing his network needs different levels of granularity for his operations. On one hand, calculating paths and installing services is preferably executed on a network-wide view. On the other hand, device specific settings and monitoring need fine granular node view with access to each device. For a routing and spectrum assignment (RSA), a combination of both might be needed to identify the internal node constraints as well as the network constraints.

From an SDN controller's point of view, the first and most important task is to establish a topology. This means discovering the devices and links of the managed infrastructure. It is a prerequisite for many tasks that rely on a network scope. One of them is path computation, which is performed on a topology representation of the network. This representation needs to be aware of constraints. In many cases, the computation is done by a dedicated path computation element (PCE), but it can also be part of a controller interface model. In a subsequent step, the computed paths are typically used to establish services or to optimize current assignments. Preferably, a service setup is executed on a network scope, defining the endpoints and some constraints, similar to an NMS. Nevertheless, it can be done on a per device basis, leading to an increased complexity. The configuration of individual devices is also important for provisioning new equipment, changing fiber maps, or defining port capabilities. In addition, monitoring is a major task for operations. On a device level, optical characteristics can be captured, for example, signal properties, and evaluated. If a controller or an NMS aggregates monitoring data from different sources, additional information and metrics with a network scope can be calculated, for example, the location of a fiber break by correlating alarms.

The virtualization of networks is a field of growing interest. Virtual networks need to be mapped to their physical counterparts and exposed to the client. Likewise, (virtual) configuration and service requests need to be translated to hardware directives. An intuitive way of requesting virtual networks could be realized by an intent interface. An intent specifies what the user wants to achieve without the need of being aware of how it is done. Even though we focus on optical networks, we do not want to ignore other network layers at the edge and briefly consider transitions, which are needed for efficient multilayer operations. Finally, network management should be able to adapt to new technologies. One example for a current development is the standardized flexible DWDM grids, which require additional parameters for the spectrum assignment. The network management functions and their assignment to scopes are summarized in Table 1.

Table 1: Assignment of network management functions to scopes.

| | |
|---|---|
| Network scope | Topology discovery, path computation, connectivity management, network monitoring, topology virtualization, multilayer operations |
| Node scope | Inventory discovery, device configuration, device monitoring |

## 4. Available Models

We will investigate four selected models, which currently receive the most attention among vendors, providers, and operators. The models are ONF Transport API (TAPI), IETF TE topology, OpenConfig, and OpenROADM.

*4.1. ONF TAPI.* The Open Networking Foundation (ONF) defines an application programming interface (API) for transport networks, called Transport API [10]. The ONF core information model [8], which describes a generic network model, serves as a starting point. The TAPI exposes information that is relevant for applications and controllers.

The topology is defined as a set of network resources which consists of nodes and links, including extended logical TPs from the core model (see Figure 1). Three logical TP types are defined in the TAPI. Node-edge-points cover the inward facing aspects, that is, forwarding capabilities. Being the entry and exit points of a node, the endpoints are responsible for encapsulation. Service-end-points are concerned with the outward facing aspects. They provide an abstracted view to the client and can map to multiple node-edge-points. The last type is connection-end-points. They cover ingress and egress aspects of connections, which represent enabled forwarding capabilities, and have a client-server relationship with node-edge-points.

The TAPI covers a set of five control plane functions and services. The *topology service* is concerned with the retrieval of information about topologies. Different granularities are defined, ranging from the whole topology down to individual node-edge-point details. The *connectivity service* is used to create and manage services between service-end-points. Connectivity is realized by underlying connections that represent forwarding behavior and are associated with connection-end-points. Connections can be built recursively and contain routes. At higher layers, a route is a list of connections in the underlying topology. At the lowest layer, they correspond to a switch matrix. The service interface does not require prior knowledge of the topology. The *path computation service's* main purpose is to compute and optimize point-to-point paths. The *virtual network service* allows the user to request virtual network topologies based on traffic constraints between pairs of service-end-points. They are implemented by reserved resources that can be controlled by the client. Finally, the *notification service* is one of the latest additions to the model. The notification types are currently limited to created and deleted objects and changed parameter values and states. Modelling of alarms and performance monitoring is scheduled for the next release.

*4.2. IETF TE Topology.* The TE (Traffic Engineering) topology is presented following the hierarchical import dependency summarized in Figure 2.

*4.2.1. Network and Network Topology.* The *IETF-network* [11] describes simple network hierarchies and their relation to each other. A network consists just of nodes and supporting networks. A network may have supporting networks corresponding to underlay networks. The nodes are defined relative to a network and cannot be reused for multiple networks. Nodes may have supporting nodes, which are part of a supporting network.

The *IETF-network-topology* defined in the same draft [11] extends the pure network model by adding termination points to the nodes and links that are interconnecting them. Therefore, topologies are supported, while the previously defined hierarchies are preserved. Termination points and links may have supporting entities of the same type. Keeping the generic nature, there are no constraints on the particular implementation, for example, physical or logical port.

*4.2.2. TE Topology.* The *IETF-TE-topology* [12] describes one way of storing traffic engineering data in a TE database (TED). The representation supports path computation that is aware of physical constraints while remaining protocol-independent. The vanilla TE Topology defines a generic view of a given network that is discovered using any available technique or protocol. In general, the TE Topology is the control plane representation of the data plane topology. A topology provider can create a virtual overlay topology for a client, which is called abstract topology in this context.

The TE Topology comprises TE links and nodes. A TE node represents a fraction of one or multiple underlying nodes and is bound to a particular topology. Multilayer TE nodes can be decomposed into a client and server layer. A transition is represented by a link between those layer nodes. A connectivity matrix shows valid interconnections between link TPs. A calculated TE path describes the nodes and links of a potential connection. TE Topologies can build hierarchies in which an underlay topology serves as a base for the overlay topology.

*4.2.3. Technology Specific Extensions.* To implement a network representation, the TE Topology needs to be augmented with technology specific details.

The *IETF-WSON-topology* [13] adds a description of impairment-unaware WSONs (Wavelength Switched Optical

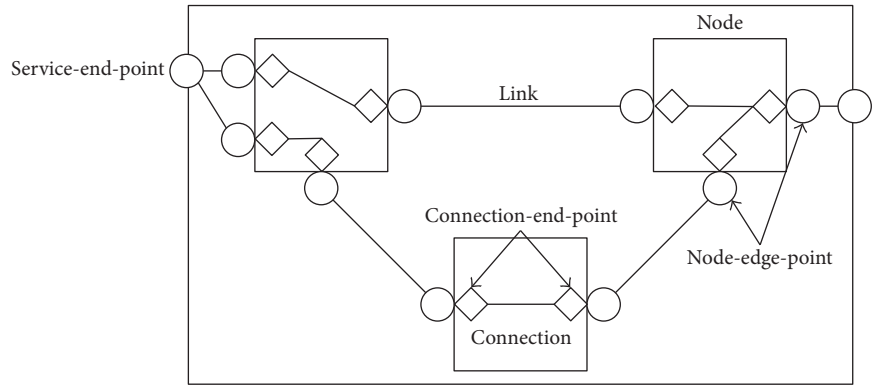

Figure 1: TAPI topology components.

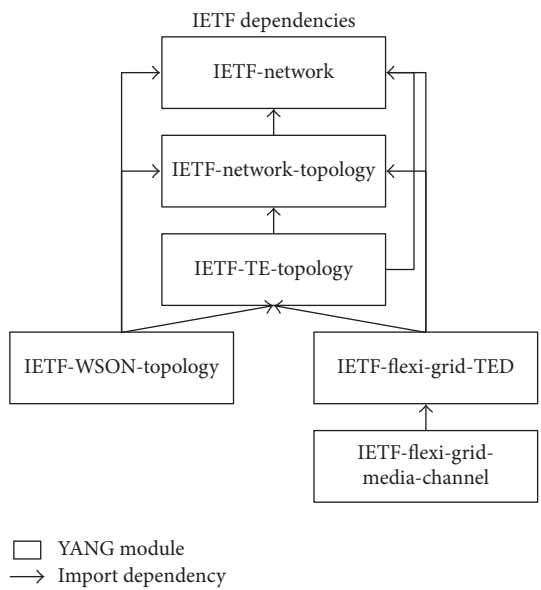

Figure 2: Import hierarchy of IETF YANG models.

Networks). The goal is to support a routing and wavelength assignment. The important extensions are node types for fixed and reconfigurable optical nodes and the augmented WSON connectivity matrix for describing available cross connections.

The flexi-grid extension *(IETF-flexi-grid-TED/IETF-flexi-media-channel)* is defined in [14]. It comprises two definitions: a TED definition for flexi-grid equipment and a media-channel description for paths. The TED part defines three types of flexi-grid nodes: node, transponder, and sliceable transponder. The node represents a wavelength switch based on a connectivity matrix. The transponders define the transmission parameters, for example, modulation format. The actual grid information is stored inside the links. The separate media-channel description extends the TED representation by a path definition including the spectrum assignment.

*4.3. OpenConfig.* OpenConfig (http://www.openconfig.net/ [accessed on 31.03.2017]) is a project by network operators to define model-driven network management interfaces, which are vendor-independent. The model tries to capture configuration and state parameters. Additionally, performance monitoring is a central part of the model. We only consider the five models related to optical transport (version 0.4.0).

The *openconfig-transport-type* model is a collection of types and identities for optical devices. The first common data elements are defined in *openconfig-transport-line-common*. Optical line ports are assigned a certain type, that is, in, out, add, or drop. The optical supervisory channel (OSC) configuration captures the available OSC interfaces.

The *openconfig-optical-amplifier* model describes optical amplifiers that are deployed as part of the transport line system, for example, EDFA and Raman. The gain parameters, output power and mode of operation, can be configured and the actual gain values and the input power are captured.

Terminal systems—client and line side—are defined in *openconfig-terminal-device*. The terminal system description is following the client to line direction. The opposite direction is implicit. The physical port represents a physical, pluggable client port on the device with operational state and performance monitoring. Each physical port has one or more physical channels. Their main purpose is to allow individual monitoring of channels that build up the full port capacity. From a model perspective, the logical channel ingress defines the contributing transceiver and physical channels. Logical channels are a grouping used for representing logical grooming elements. They are assigned either to another logical channel, in order to add another stage of multiplexing (or demultiplexing), or to an optical channel, which corresponds to the line side transmission and assigns a carrier with a wavelength or frequency. The two defined protocol types for logical channels are Ethernet and optical transport network (OTN). In general, the model assumes that the NMS will verify correct combinations of protocols.

Finally, the *openconfig-wavelength-router* model contains a definition for optical transport line system nodes or ROADMs (Reconfigurable Optical Add-Drop Multiplexers). A wavelength router is defined as a configurable switching element. A media channel is described by a lower and upper frequency, therefore specifying a frequency band. This media channel is then assigned an input and output port, thereby defining the flow inside of the node.

*4.4. OpenROADM.* OpenROADM is an initiative defining a white-box model for ROADM based optical equipment that can be used by a control plane, that is, OpenROADM controllers [15]. It creates a network view that abstracts vendor-specific devices to a generic device representation, which can be used for service instantiation or path computation. The modelling is closely coupled to the structure of a ROADM node, leading to two basic building blocks, direction/degree and add/drop group. A direction or degree is the block that is connected to the degree of another ROADM node on one side and the internal structure on the other side. Add/drop groups are a unit that allows adding and dropping wavelengths between fibers and client ports. In the model, those client ports are grouped together based on shared risk groups (SRGs). The connected client equipment, including transponders and routers, is called tail.

Degrees have two types of (logical) TPs: trail and connection TPs. The trail TPs are facing a degree of an adjacent ROADM and terminate optical multiplex section (OMS) trails. Connection TPs (CTPs), on the other hand, are used to connect add/drop ports and pass express traffic inside of a node.

An add/drop group is a construct consisting of wavelength selective switches (WSSs), amplifiers, and splitters/combiners for transmitting and receiving signals. Transponders are connected to add/drop ports and connection points (CP) are facing the degrees. Add/drop groups are assigned one or more SRG depending on the hardware configuration. Every SRG contains one pair of CPs. In addition, various alarms and performance measurements are available at this level.

Logical connectivity links are a tool to represent the connectivity/cabling between building blocks. They cover external connectivity between degrees and the internal cabling between degrees and SRGs. Express links are used for transit traffic, internally passing from one degree to another (CTP ports). The connectivity map is based on input from planning. Wavelengths are represented as fan-out at the connection points. Each wavelength receives an own node. It multiplies the number of termination points by the number of available wavelengths per direction.

The path computation for a service can be performed between either tails or nodes. In the second case, appropriate SRGs and transponders (if available) are returned. The services are purely wavelength-based ignoring any other parameters. Only routing constraints may be applied to influence the outcome of the path computation, for example, diversity, inclusion, and exclusion.

## 5. Evaluation of Models for Optical Networks

The models are evaluated based on the previously presented management functions. The evaluation is summarized in Table 2.

We start with the classification of the models according to [16]. The TAPI and TE Topology are standard models for network services. OpenROADM and OpenConfig, on the other hand, are models at the network element level and are vendor-specific according to the definition, even though OpenROADM can be seen as having a network scope to some extent; three of the surveyed models are able to represent topologies. Only OpenConfig is lacking the notion of links or equivalent entities. This makes it unsuitable for path computation on its own. While the TAPI provides an application interface for retrieving paths, TE Topology and OpenROADM are representations that can be used to perform the actual computation. The calculated paths can be installed through the connectivity service of the TAPI or the service remote procedure calls of OpenROADM. The TE Topology can only be used in combination with an active stateful PCE to manage services. OpenConfig is not capable of such network-wide operations without an additional entity with knowledge about the topology. However, both device-oriented models, that is, OpenConfig and OpenROADM, are able to configure individual network elements. The TE Topology is limited to the information available for the path computation, whereas the TAPI focuses on network-wide operations and the individual devices are out of scope. Notifications and performance monitoring are management tasks that the network element scoped protocols support very well. Being intended for a TED representation, the TE Topology does provide notification mechanisms about changes. To some extent, the same is true for the TAPI and further extensions towards alarms and performance monitoring are planned. Virtual topologies are only supported by the TAPI and TE Topology. At the edge of the optical network, interworking with other layers is required. The TAPI defines a set of known signals, for example, OTN and Ethernet. The generic TE Topology requires extensions to support any technology specifics but is very flexible in using them. OpenConfig comprises models for many protocols, like IP, Ethernet, and MPLS. The interworking takes place at an individual device level. Finally, OpenROADM supports transponders and pluggables at its edge. For external devices, additional information needs to be handed over to the controller.

The final point, which we investigate, is the consideration of current developments. We pick flexible DWDM grids as an example that has been standardized but is not widely deployed yet. Most of the models do not provide any information on the applicability to flexible grid scenarios. We therefore evaluate the current definition and the needed adjustments. In the TAPI, a very generic definition of a central frequency or wavelength is provided. This is not enough to describe flexible grids but it can be a starting point for augmentation. For the TE Topology, a specialized extension is already available. In the case of OpenConfig, one frequency value per optical channel is available. Therefore, it is not capable of representing the new grids without changes. OpenROADM is focused on capturing the current generation of ROADMs. Particularly the representation of wavelengths by individual connection points makes an adaptation very hard without major changes of the model.

All presented models have a particular application area in mind. To support network virtualization, we mapped the models to generic optical network hypervisor architecture (Figure 3 rhs). The hypervisor takes the role of a mediator between the client's SDN controller and the network elements. Starting from the top, the TAPI covers a wide range of interfaces needed for applications and controllers. This means that it is a good fit for the northbound API of the hypervisor, which requires support of network-wide operations. In general, the IETF-TE-topology is an extendable representation of networks for TEDs. This makes it applicable to the internal representation of the hypervisor. The overlays are a good fit for creating virtual topologies for clients. The functionality that is usually added by an attached PCE is taken over by the hypervisor. For the configuration of the network elements, that is, the southbound interface, OpenConfig and OpenROADM are possible solutions. The optical part of OpenConfig is still in an early stage and focuses on the configuration and monitoring of individual components of an optical transport line system. Many parts of the optical model are currently undefined. OpenROADM provides a device as well as a network view. It is very operations-oriented, meaning that all components have associated management parameters like physical location and vendor information.

Table 2: Evaluation of the selected optical models.

| | TAPI | TE topology | OpenConfig | OpenROADM |
|---|---|---|---|---|
| Classification [16] | Network/standard | Network/standard | Device (+network)/vendor | Device/vendor |
| Topology representation | Yes | Yes | No | Yes |
| Path computation interface | Yes | No (TED for PCE) | No | No (TED for PCE) |
| Network wide operations (e.g., services) | Yes | Only with active PCE | No | Yes |
| Explicit device configuration | No | High level | Yes | Yes |
| Notifications and performance monitoring | Notifications (alarms and PM planned) | Notifications | Yes | Yes |
| Network virtualization | Yes | Yes | No | No |
| Interworking with other layers | Predefined set | Through extensions | On a device basis | Transponders and pluggables |
| Flexible DWDM grids | No | Yes (extension) | No | No |

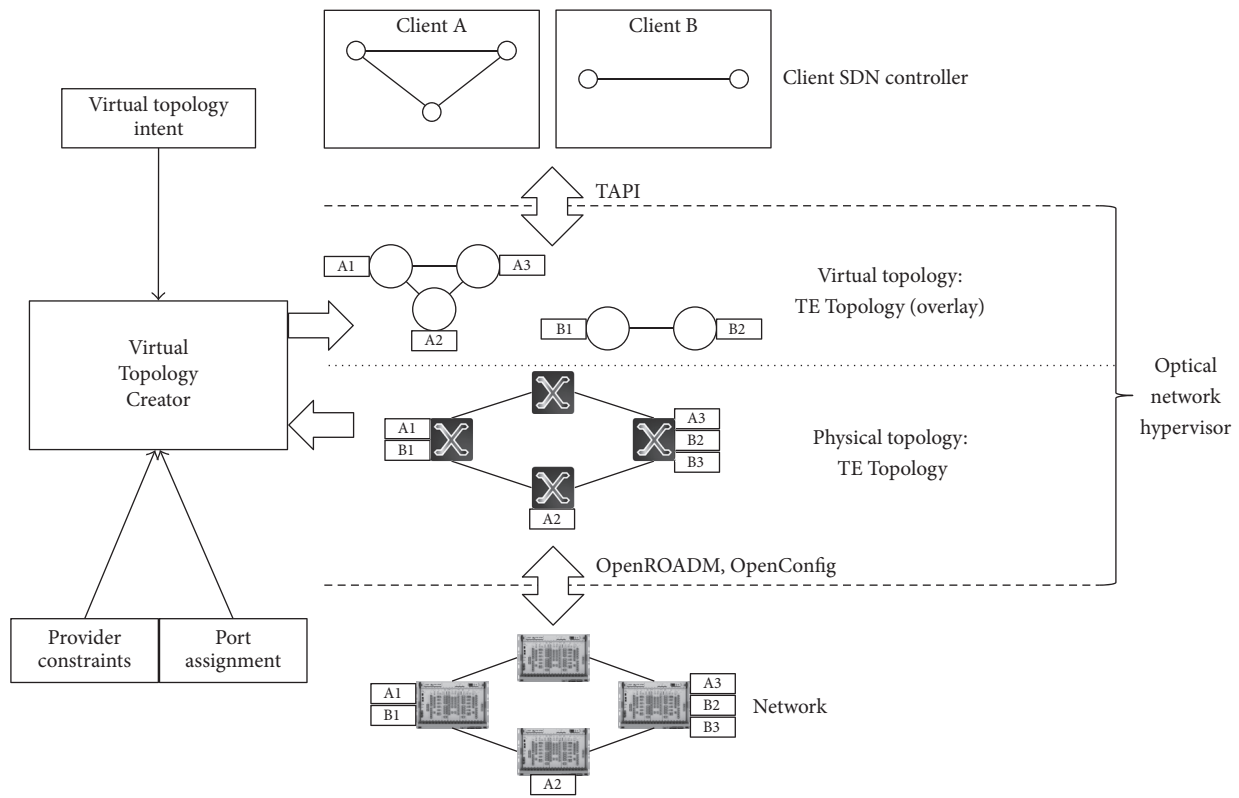

Figure 3: Generic optical network hypervisor architecture with a virtual topology intent interface (lhs) and the model assignment (rhs).

Two of the presented models, that is, TAPI and TE Topology, allow a creation of virtual topologies. In [17, 18], defining interfaces for network virtualization was identified as one key research challenge. So far, there is still no straightforward way for the customer or an application to define a topology by simply indicating his or her intent. Compared to network embedding, creating a virtual topology for optical networks, which can then be controlled by a client SDN controller, does not include an assignment of nodes in the network. The endpoints for potential connections are allocated to the customer by the infrastructure provider beforehand. The main task is to create links based on impairment-aware RSA in order to receive a topology.

## 6. Intent Model for Defining a Virtual Network Topology

The creation of virtual topologies is a base technology and enables the virtualization and slicing of the optical transport network for multiple tenants and scenarios, for example, network slicing for 5G. Current work [19] mentions two ways to create a virtual topology: the first one is a direct definition in one of the topology models and the other one is based on a traffic matrix, whose structure has not been defined yet. The first approach is supported by the TE Topology and the TAPI follows an approach similar to the second one. Reference [20] suggests an intent-based network definition using a graph. Endpoints are grouped in policy domains (nodes) and connected via policies (links). No details about the available parameters, the assignment of nodes, and an exchange model have been published. In [21], three groups of virtual topology intents are introduced in the sense of high-level policies. A topology intent describes a connectivity between endpoints. Endpoint intents capture the requirements for relations between endpoints, like bandwidth. Finally, chain intents extend the previous one by chaining network functions. Unfortunately, no details on the intents or their modelling are given and the proposed architecture is only applicable to switches.

Current solutions for creating virtual topologies, like entering the network layout manually, are overly complex and error-prone and most of all cannot be easily defined and modified by the client. The customer should be able to express his topology request based on a set of simple intents. We consider this interface a missing piece in software-defined optical network management and propose an intent interface model to solve it. The resulting topology can then be represented, exchanged, and controlled based on existing optical models like the TE Topology and the TAPI. The basic inputs for the virtual topology creation (Figure 3, lhs) are the assigned client ports, provider constraints, and the client's intents, provided through an application programming interface. The provider constraints are out of the scope of this work but they may define the level of abstraction by adjusting the visibility of network details.

Our proposed model enables the user to define virtual topologies expressing his requirements in the form of intents and it is defined in the modelling language YANG. The model comprises two main data entry points: `endpoints` and `topologies`. The `endpoints` path represents the state that cannot be changed by the user. It contains a list of all endpoints assigned to the client who is querying the interface. It is important to give the client a way to access this information as a starting point for his virtual topology requests. The assignment will be maintained by the provider based on existing agreements. The endpoints themselves are generic and can represent nodes, modules, or ports, depending on the level of abstraction that the interface applies. Additional information can be added by means of augmentation, for example, user-friendly descriptions. From the point of view of virtual topology creation, only a unique endpoint identifier is needed.

The configuration and therefore the creation are carried out through the data stored in the `topologies` subtree. This top-level element contains a list of installed topologies for this particular user. Those topologies comprise an identifier as well as a list of intents. The uniqueness of the topology identifier is enforced by the key property for the list, which is also true for the other identifiers. The individual intents combine a set of endpoints with requirements that need to be fulfilled by these endpoints. Those parameters include the dedicated bandwidth, which is exclusively reserved, and the flexible bandwidth, which is shared with others and might not always be available. Both are expressed in `Mbit/s`, thereby giving the user an easy way to define a topology based on a bitrate instead of expecting him to be familiar with technology-dependent details, like flexible WDM grids and modulation formats. Additionally, parameters addressing properties of the available paths, that is, links in the virtual topology, are included. Those comprise the minimum number of parallel paths and their disjointness requirements, for example, link or node disjoint. It is also possible to request an optical protection. Finally, the maximum number of parallel connections is included. It defines how many connections may be active at any given point in time. This means that the links define potential connections but only a certain number of them can be provisioned by the user at the same time. Based on a number of one or more such intent groups, a topology is created and exposed to the client's SDN controller, which is then able to control the virtual network on demand. An example for assigned endpoints and a topology based on one intent is shown in Algorithm 1. If our YANG model was used to define a RESTCONF interface, the JSON representation would match a response to a request for the current configuration of the virtual topologies. The example corresponds to client A's topology in Figure 3.

The next step will be defining an algorithm for the Virtual Topology Creator (Figure 3) that is capable of collecting the individual intents, combining them, and creating a virtual topology. For the creation of the virtual topology, technology details and physical constraints need to be taken into account and network embedding strategies, which are out of scope for the work, can be applied. The focus is on creating a workflow that is capable of starting from intents and providing a virtual topology, which maps to a part of the physical network. To this means, implementation of the full architecture shown in Figure 3 is needed to be able to evaluate the approach experimentally. In a subsequent step, this base model can be augmented to support network slicing by adding additional parameters, like availability or latency, as well as provisioning the virtual network right away.

```
{
  "endpoints":{
    "assigned-endpoints":[
      {"endpoint-id":"A1"},
      {"endpoint-id":"A2"},
      {"endpoint-id":"A3"}
    ]
  },
  "topologies":{
    "installed-topologies":[
      {
        "topology-id":"Client A",
        "intents":[
          {
            "intent-id":"Intent A",
            "endpoints":["A1", "A2", "A3"],
            "dedicated-bandwidth":10000,
            "flexible-bandwidth":5000,
            "minimum-paths":2,
            "disjoint-paths":"link",
            "protection":false,
            "maximum-active-connections":2
          }
        ]
      }
    ]
  }
}
```

Algorithm 1: JSON representation of an example for a virtual topology configuration for client A corresponding to the depicted virtual topology in Figure 3.

## 7. Conclusion

In this paper, we surveyed optical models for software-defined network management. We have briefly introduced and evaluated the most popular models for optical networks, that is, ONF Transport API, IETF-TE-topology, OpenConfig, and OpenROADM. We have explained their respective strengths and weaknesses and have shown their application areas by mapping them to components of a generic optical network hypervisor. By doing so, we identified a gap between the usage and definition of virtual topologies and proposed an intent interface model to close it. The intent interface takes a set of requirements by a user, translates them to a virtual topology, and hands over the control to the user's SDN controller. Based on this interface, software and algorithms for creating virtual topologies can be developed.

## Conflicts of Interest

The authors declare that they have no conflicts of interest.

## Acknowledgments

Parts of this work have received funding from the European Union's Horizon 2020 Research and Innovation Programme under Grant Agreement no. 645127 (ACINO) and from the European Research Council (ERC) under the auspices of the European Union's Horizon 2020 Research and Innovation Programme (Grant Agreement no. 647158—FlexNets).

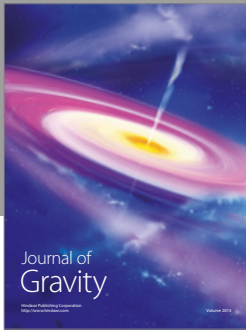
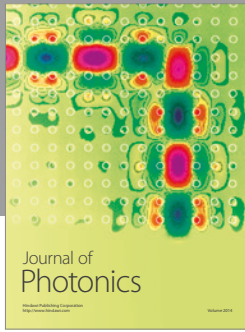
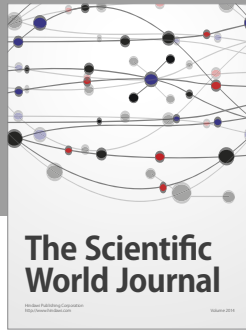
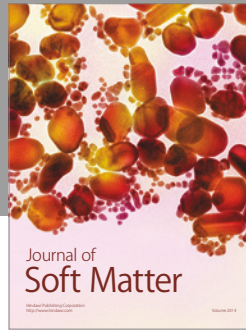
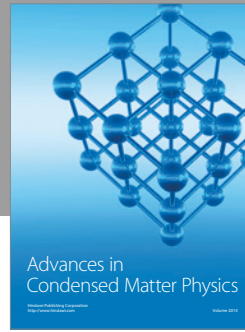
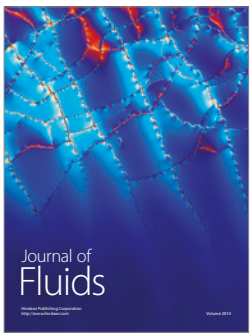
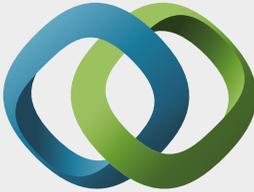
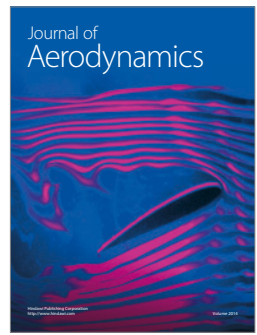
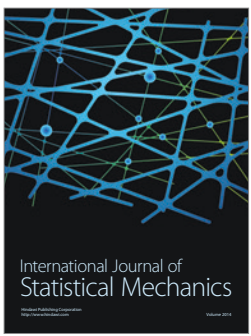
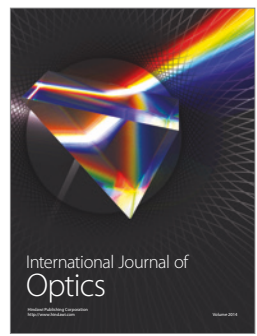
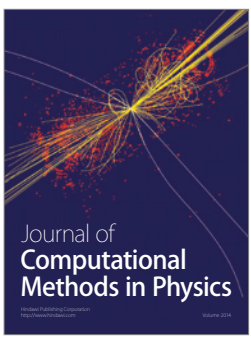
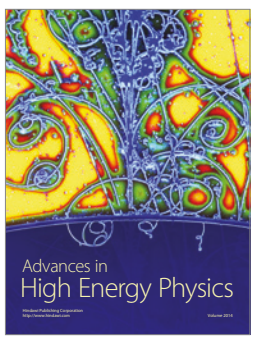
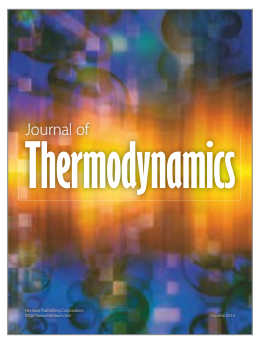
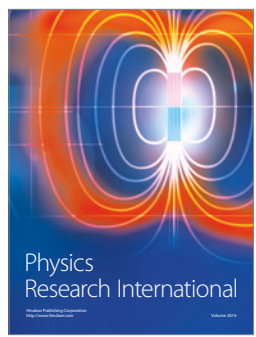
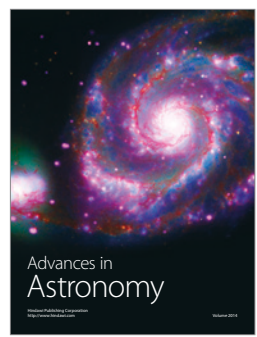
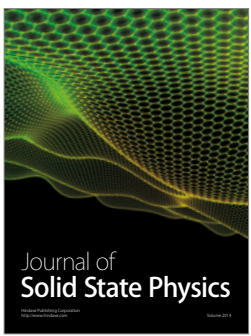
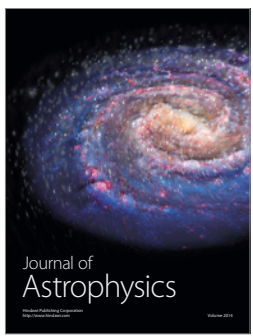
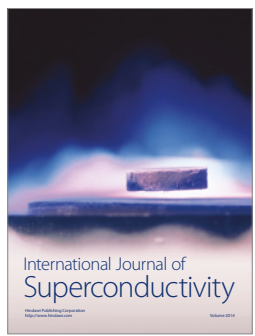
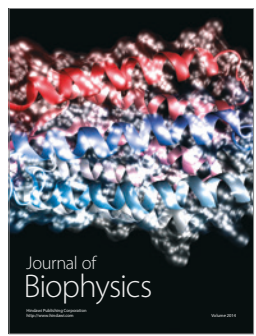
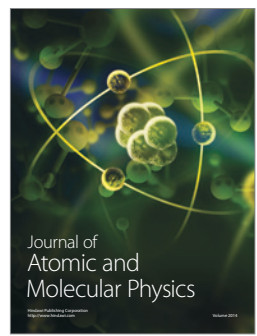